\documentclass[nofootinbib,aps,11pt]{revtex4}
\usepackage[english]{babel}
\usepackage{amsmath}
\usepackage{amssymb}
\usepackage{graphicx}% Include figure files
\usepackage{dcolumn}% Align table columns on decimal point
\usepackage{bm}% bold math
\usepackage{textcomp}
\usepackage{wasysym}
\usepackage{stmaryrd}
\usepackage{slashed}

\usepackage{pst-all}

\def \ee{\end{equation}}
\def \be{\begin{equation}}

\begin{document}

\title{\Large  Renormalization group improved black hole
space-time in large extra dimensions}
\author{Thomas Burschil$^*$}
\author{Benjamin Koch$^\dag$}
\affiliation{
{\small 
$^*$Institut f\"ur Theoretische Physik, Johann Wolfgang Goethe -
Universit\"{a}t,
Max-von-Laue-Strasse 1, D-60438 Frankfurt am Main, Germany;\\ 
$^\dag$Departamento de F\'\i sica, Pontificia Universidad Cat\'olica 
de Chile, Avenida Vicu\~na Mackenna 4860, Santiago, Chile.
}}
%%%%%%%%%%%%%%%%%%%%%%%%%%%%%%%%%%%%%%%%%%%%%%%%%%%%%%%%%%%%%%%%%%%%
\begin{abstract}
By taking into account a running of the gravitational
coupling constant with an ultra violet fixed point, an improvement
of classical black hole space-times in extra dimensions is studied.
It is found that the thermodynamic properties 
in this framework allow for an effective 
description of the black hole evaporation process.
Phenomenological consequences
of this approach are discussed and the LHC discovery
potential is estimated.\end{abstract}
%%%%%%%%%%%%%%%%%%%%%%%%%%%%%%%%%%%%%%%%%%%%%%%%%%%%%%%%%%%%%%%%%%%%
\date{\today}
\maketitle

%%%%%%%%%%%%%%%%%%%%%%%%%%%%%%%%%%%%%%%%%%%%%%%%%%%%%%%%%%%%%%%%%%%%%%%%%%%%5
\section{Introduction}

Models with extra spatial dimensions offer an elegant
solution to the hierarchy problem \cite{ArkaniHamed:1998rs,Randall:1999vf}. 
In the case of \cite{ArkaniHamed:1998rs} this is achieved
by banning all standard model particles and forces onto a 4-dimensional
subspace, while gravity can propagate also into $d$ additional
spatial dimensions. In order to keep the model consistent
with todays gravity experiments the additional dimensions
are assumed to be compactified in a small volume $V_d$.
By this construction the 
measured gravitational coupling (or equivalently the
Planck mass $M_{Pl}$) can be explained by a 
fundamental mass $M_f$ which might be as low as a few TeV. 
Since this would be much closer to the electro-weak scale,
such models give a possible solution of the hierarchy problem.
The relation
\begin{align}
   M^2_{Pl} &= V_d M_f^{d+2}
   \label{eq:int:relation}
\end{align}
connects these two couplings via the volume $V_d$ which is spanned by the extra
dimensions . 
In a world with extra dimensions and a gravitational coupling in a TeV range
colliders like the LHC could create tiny black holes (BH)
\cite{Banks:1999gd,Giddings:2000ay,Giddings:2001bu,
Dimopoulos:2001hw,Hossenfelder:2001dn,Bleicher:2001kh}. 
The line element of a higher dimensional spherically symmetric black hole
is given by \cite{Myers:1986un}
\begin{align}
   ds^2 &= f(r) dt^2 - f^{-1}(r) dr^2 - r^2d\Omega_{d+2}
   \label{eq:int:line}\\
{\mbox{with}}\quad   f(r) &= 1 - \frac{R_H^{d+1}}{r^{d+1}} \ .
   \label{eq:int:koeff}
\end{align}
The event horizon $R_H$ depends on the black hole mass $M$ and 
the universal gravitational coupling $G$
\begin{align}
   R_H^{d+1} &= \frac{16\pi GM}{(d+2)A_{d+2}}\ ,
   \nonumber\\
\mbox{where} \qquad A_{d+2} &=
\frac{2\pi^{\frac{d+3}2}}{\Gamma\left(\frac{d+3}2\right)}
\nonumber
\end{align}
marks the surface of a $d+3$ unit sphere. 
%\footnote{%
Please note, that there are different
definitions of the higher dimensional coupling constant $G$. We use the
definition of \cite{Dimopoulos:2001hw}. For more discussion and other
definitions see \cite{Giddings:2001bu}.
%}
By redefining the coupling in
terms of the fundamental mass
$M_f^{d+2} \equiv 1/G$, the radius of the event horizon is given by
\begin{align}
   R_H^{d+1} &= \frac{16\pi}{(d+2)A_{d+2}} \frac {M}{M_f^{d+2}} \ .
   \label{eq:int:rh}
\end{align}
This form of the black hole horizon holds as long 
as $R_H \ll R$ which is for TeV masses true since
$R_H$ exceeds $R$ typically by fifteen orders of magnitude.
A black hole emits thermal radiation 
\cite{Hawking:1974rv}. 
The temperature of this radiation is given by the
radial derivative of the metric coefficient $f(r)$
at the horizon.
For the case of $d$ extra dimensions this temperature is given by
\begin{align}
   T_{H} = \frac{d+1}{4\pi R_H} \ .
   \label{eq:TH}
\end{align}
However, this prediction is limited to large black hole
masses $M\gg M_f$.
For masses close to the fundamental mass 
one expects modifications of the Hawking temperature
and it was conjectured that the thermal radiation could be
suppressed, leading to the formation of a stable
final state \cite{relics,Koch:2005ks,Vilkovisky:2005cz}. \\

%\subsection{Renormalization group}
Although, some extra dimensional models
like in eq.\;(\ref{eq:int:relation}) can solve the 
hierarchy problem, they are not the desired unified
description of all forces yet. 
The reason for this is that
 gravity (with or without extra dimensions)
can not be generalized in the usual loop expansion to a renormalizable quantum
field theory. 
It was conjectured that this problem results from
expanding the theory in the gravitational coupling instead
of solving the complete theory that might even contain
higher powers in the curvature $R$.
In \cite{Reuter:1996cp,Litim:2003vp,Niedermaier:2006ns,Litim:2008tt} 
it was shown that by using
special truncation methods an exact renormalization group (RG)
equation for the gravitational coupling can be derived.
In a first order truncation  
those studies have been generalized to extra dimensions
\cite{Fischer:2006fz,Hewett:2007st}
leading to a fundamental mass that depends on the energy scale $k$:
$M_f^{d+2} \to \tilde M_f^{d+2}(k)$.
It was shown that the running gravitational coupling has the form 
\begin{align}
   \tilde M_f^{d+2}(k) &= M_f^{d+2}\left[ 1 + \left(\frac{k}{tM_f}\right)^{d+2}\right]
   \label{eq:run:coupling}
\end{align}
which also depends on a parameter, $t$. 
Going beyond one loop, this transition behavior
between the infrared and ultraviolet regime was
found to be even more pronounced with increasing
number of extra dimensions \cite{Litim:2007iu}. 
For the case of $d=0$ the effect
of this running coupling on the structure
of the Schwarzschild metric was derived in \cite{Bonanno:2000ep}.
The aim of this paper is to repeat the construction for
$d\neq 0$ and to study its phenomenological implications.

%%%%%%%%%%%%%%%%%%%%%%%%%%%%%%%%%%%%%%%%%%%%%%%%%%%%%%%%%%%%%%%%%%%%%%%%%%%%5
\section{RG improved black holes in extra dimensions and black hole remnants}

In flat space-time the de Broglie relation connects energies 
$k$ and distances $d$ by $k=1/d$.
In curved space-time more care is needed since distances are determined
locally by the metric. For the case of a
spherically symmetric Schwarzschild space-time, modifications
of the de Broglie relation can only depend on the radial coordinate $r$.
This leads to the ansatz
\begin{align}
   k(r) &= \frac \xi{d(r)} \ ,
   \label{eq:ident:scale}
\end{align}
where $\xi$ is a parameter of order one.
Before calculating the distance function $d(r)$
it is essential to remember its behavior for
large distances $r\to \infty$. 
In this limit the metric should approach the flat Minkowski
metric
\begin{align}
   \lim_{r\to\infty} \frac{d(r)}{r} &= 1 \ .
   \label{eq:ident:limit}
\end{align}
In that case the scale approaches asymptotically $k(r\to \infty) \approx
\xi/r$.
The distance function is calculated via the definition of distance in
general relativity by integrating the line element
\begin{align}
   d(P) &= \int_{\mathcal C} \sqrt{|ds_{class}^2|}  
   \label{eq:ident:def}
\end{align}
along a curve $\mathcal C$. The subscript
$class$ indicates that the line element is calculated
with a fixed coupling $M_f$. We
parameterize $\mathcal C$ in Schwarzschild space-time and calculate the
distance along the curve
\begin{align}
   ds_{class}^2 &= \frac 1{f_{class}(r')} dr'^2
   \nonumber\\
   d(r) &= \int_{0}^{r(P)} \frac 1{\sqrt{|f_{class}(r')|}} dr' \ .
   \nonumber
\end{align}
The parameterization along the radial
coordinate $r'$ in the range $r' \in [0,r(P)]$ is chosen like in
\cite{Bonanno:2000ep}. In eq.\;\eqref{eq:ident:def} it is necessary to take the
absolute value of $ds$ in order to have always a positive distance. 
Due to the absolute value the distance function differs inside and outside 
of the event horizon.
Together with definition of the event horizon
\eqref{eq:int:rh} the distance function is expressed in the two regions
by
\begin{align}
   d_{r<R_H }(r) &= \int dr' \sqrt{\frac{r'^{d+1}}{R_H^{d+1}-r'^{d+1}}} 
   \label{eq:ident:dcalc1}\\
   d_{r>R_H }(r) &= \int dr' \sqrt{\frac{r'^{d+1}}{r'^{d+1}-R_H^{d+1}}} \quad.
   \label{eq:ident: dcalc0}
\end{align}
It is not possible to find a general analytic solution 
for those two distance functions.
Instead, the functions are interpolated between the two limits $r\to 0$ and
$r\to \infty$.
For small $r$ the denominator of (\ref{eq:ident:dcalc1}) simplifies and the
 small $r$ limit can be integrated
\begin{align}
   d(r) &= \int_{0}^r dr' \sqrt{\frac{r'^{d+1}}{R_H^{d+1}-r'^{d+1}}} 
   \nonumber\\
   &\stackrel{r'\to 0}{=}\int_{0}^r dr' \sqrt{\frac{r'^{d+1}}{R_H^{d+1}}} 
   \nonumber\\
   &=\frac{1}{R_H^{\frac{d+1}{2}}} \frac 2{d+3} r^{\frac{d+3}2} \ .
   \label{eq:ident:smallr}
\end{align}
For large $r$ one has to integrate in two steps, first form $r'=0$ to $r'=R_H$
which just gives a constant summand $\tilde B$, and afterwards from $r'=R_H$ to 
$r'=r$. In the second integration the fraction simplifies to a constant and
distance behaves like $r$, again with a summand $\tilde A$
\begin{align}
   d(r) &= \int_{R_H}^r dr' 
\sqrt{\frac{r'^{d+1}}{r'^{d+1}-R_H^{d+1}}} 
\nonumber  \\ \nonumber
   &\stackrel{r'\to \infty}{=}\int_{R_H}^r dr'
 \sqrt{\frac{r'^{d+1}}{r'^{d+1}}}\\ 
   &= r - \tilde A \ .
  \label{grossr}
\end{align}
This is the dependency as required in \eqref{eq:ident:limit}. 
Now the distance function is interpolated between 
eq. (\ref{eq:ident:smallr}) and (\ref{grossr}) by
\begin{align}
   d'(r) &= \left(\frac{r^{d+3}}{r^{d+1} + {\gamma_d}
   R_H^{d+1}}\right)^{\frac 12} 
   \nonumber\\
   \gamma_d &= \frac{(d+3)^2}4 \ ,
   \label{eq:ident:g}
\end{align}
which has the correct asymptotic behavior in the limits 
$r \to \infty$: $d'(r)\to r$ and $r\to 0$: $d'(r) \to r^{(d+3)/2}$. 
The parameter $\gamma_d$ is evaluated by the small $r$ limit.
Identifying the energy scale $k$  with inverse distance one finds
\begin{align}
   k(r)&=\frac{\xi}{d'(r)}
   \nonumber\\
   &= \xi \left(\frac{r^{d+1} + {\gamma_d}
   R_H^{d+1}}{r^{d+3}}\right)^{\frac 12} \quad.
   \label{eq:ident:kscale}
\end{align}
This relation between the energy scale $k$ 
and the radius $r$ in higher dimensional
Schwarzschild space-time allows to express the
scale dependent fundamental mass $\tilde M_f$
\eqref{eq:run:coupling} in terms of the radial coordinate $r$. \\

\begin{figure}[hbt]
   \centering
%\centerline{\protect\vbox{\epsfig{file=fvsr_d2_tt001_Mf1.eps,
%width=0.6\textwidth}}}
\includegraphics[width=10cm]{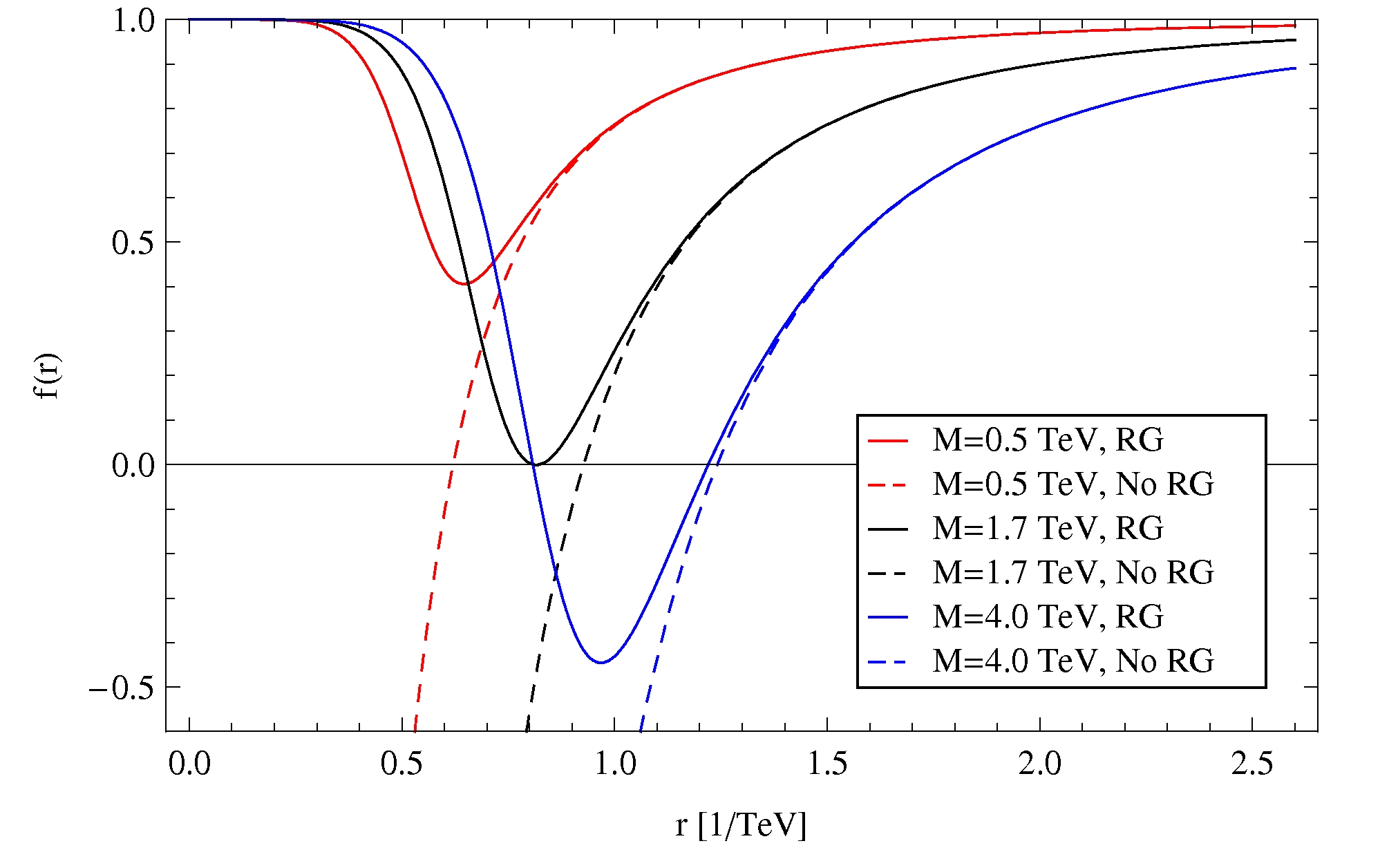}
   \caption{\label{fig:gammad2}Metric coefficients $f(r)$ for different BH
   masses $M$ at $d=2$, $\tilde t=0.002$, and $M_f=1\;{\rm TeV}$.  
  For those parameters a critical mass of
   $M_c=1.67\;\rm{TeV}$ is found.}
\end{figure}

\begin{figure}[bht]
   \centering
%\centerline{\protect\vbox{\epsfig{file=Mcritvstt_Mf1.eps,width=0.6\textwidth}}}
\includegraphics[width=10cm]{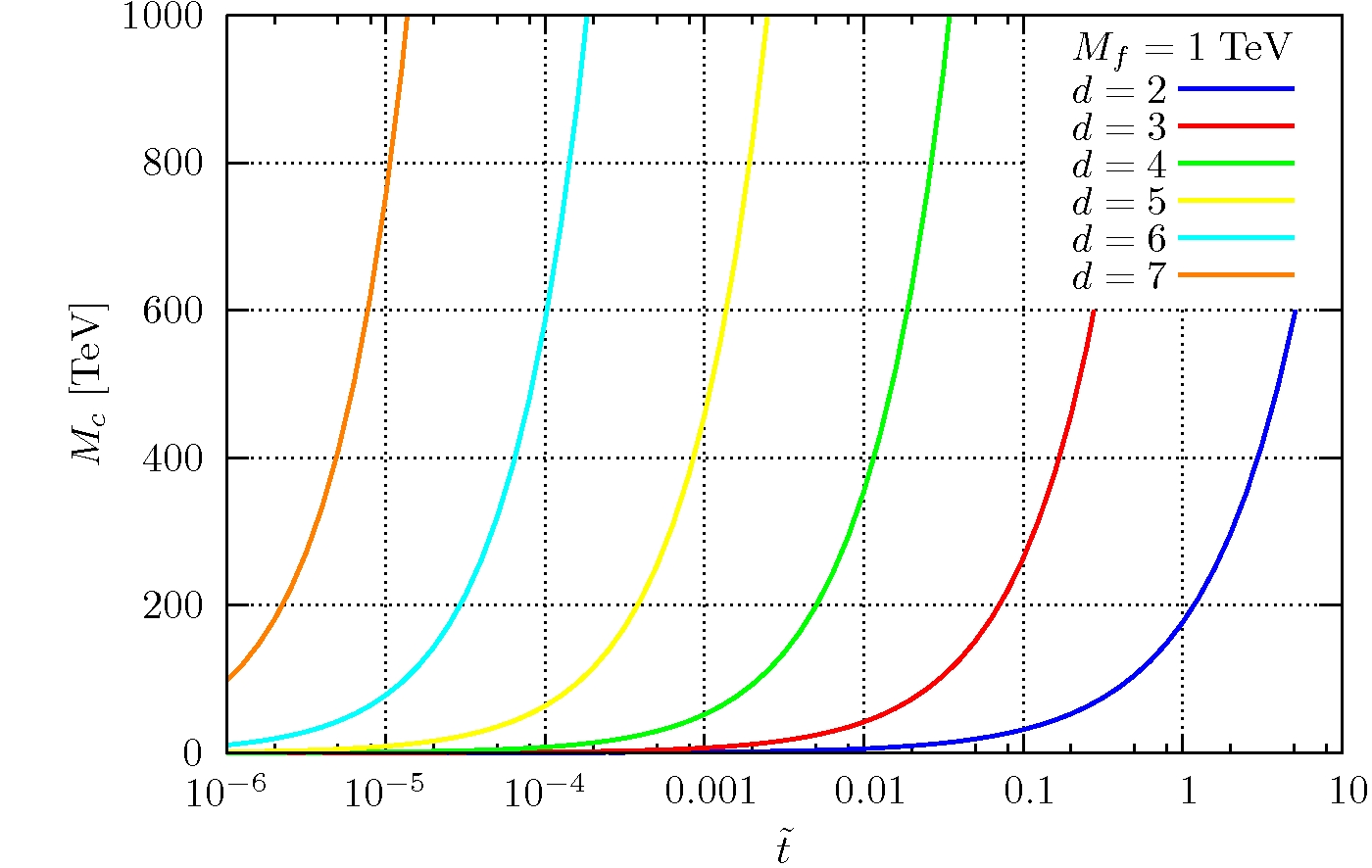}
   \caption{\label{fig:tvar}Remnant masses depending on the quantum 
gravity parameter
   $\tilde t$ for different numbers of extra dimensions $d$.}
\end{figure}

%\subsection{Schwarzschild metric and BH remnants}
Modifying the horizon radius $R_H$ by the
radius dependent fundamental mass $\tilde M_f(r)$ 
and defining
\begin{align}
   \tilde t = (\xi/t)^{d+2} \ ,
   \label{eq:imp:ttilde}
\end{align} 
gives
\begin{align}
   \tilde R_H^{d+1}(r) &= \frac{16\pi}{(d+2)A_{d+2}}  \frac
   1{\tilde M_f^{d+1}} \frac{M}{\tilde M_f}
   \nonumber\\
   &= R_H^{d+1} \left[ 1 +
   \frac{\tilde t}{M_f^{d+2}} \left(\frac{r^{d+1} + {\gamma_d}
   R_H^{d+1}}{r^{d+3}}
   \right)^{\frac{d+2}{2}} \right]^{-1} \quad, 
   \label{eq:imp:rh}
\end{align}
where $\tilde t$ parameterizes the strength of the RG
corrections to the classical result.
Since this $\tilde R_H$ has an explicit $r$-dependence
it can not be interpreted as event horizon.
Like usually the event horizon of a spherical symmetric
black hole solution is the zero of the radial metric coefficient
$f(r)=1-\tilde R_H^{d+1}(r)/r^{d+1}$.
As shown in figure\;(\ref{fig:gammad2})
%%%%%%%%%% FIGURE %%%%%%%%%%%%%%%%%%
%%%%%%%%%%%%%%%%%%%%%%%%%%%%%%%%%%%%%%%%%%%%
%
the metric function does not cross $f(r)=0$ for small values of $M$ and so there
is no singularity in the line element $ds^2$.
However, for a larger black hole mass one finds
the critical case where
the line element gets a zero at one radius.
For $M>M_{c}$ this zero splits up into two zeros,
where the outer zero corresponds to the apparent event horizon.
As also shown in figure\;(\ref{fig:gammad2}) this outer horizon grows for large BH
masses (\(r\to \infty\)) and approaches
the classical event horizon of eq.\;\eqref{eq:int:rh}. 
This kind of behavior of the metric function
is independent of the number of extra dimensions.
As it will be explained in the following sections,
it is natural to identify the critical mass with the mass of 
of a black hole remnant 
\[ M_R = M_{c} \ .\]
The critical mass can be calculated in dependence
of the parameters $\tilde t$ and $M_f$. 
As it can be seen in figure\;(\ref{fig:tvar}) for $d=2,\dots ,7$, $M_c$ depends
strongly on the parameter $\tilde t$. While $M_c$ tends
to zero for small values of $\tilde t$, it grows rapidly
beyond the experimentally testable TeV range for larger values of $\tilde t$.
%%%%%%%%%% FIGURE %%%%%%%%%%%%%%%%%%
%%%%%%%%%%%%%%%%%%%%%%%%%%%%%%%%%%%%%%%%%%%%%%%%%%%%%%%%
Since the strength of the quantum gravity corrections
is parameterized by $\tilde t$,
it is interesting to note
that by construction the continuous
limit $\tilde t \rightarrow 0$ leads to 
$M_c \rightarrow 0$, whereas in a priory classical calculation
($\tilde t=0$) no critical mass $M_c$ and therefore no remnant exists.

%%%%%%%%%%%%%%%%%%%%%%%%%%%%%%%%%%%%%%%%%%%%%%%%%%%%%%%%%%%%%%%
\section{Black hole thermodynamics}

\begin{figure}[bht]
\centering
%\centerline{\protect\vbox{\epsfig{file=TvsM_d2_Mf1.eps,
%width=0.6\textwidth}}}
\includegraphics[width=10cm]{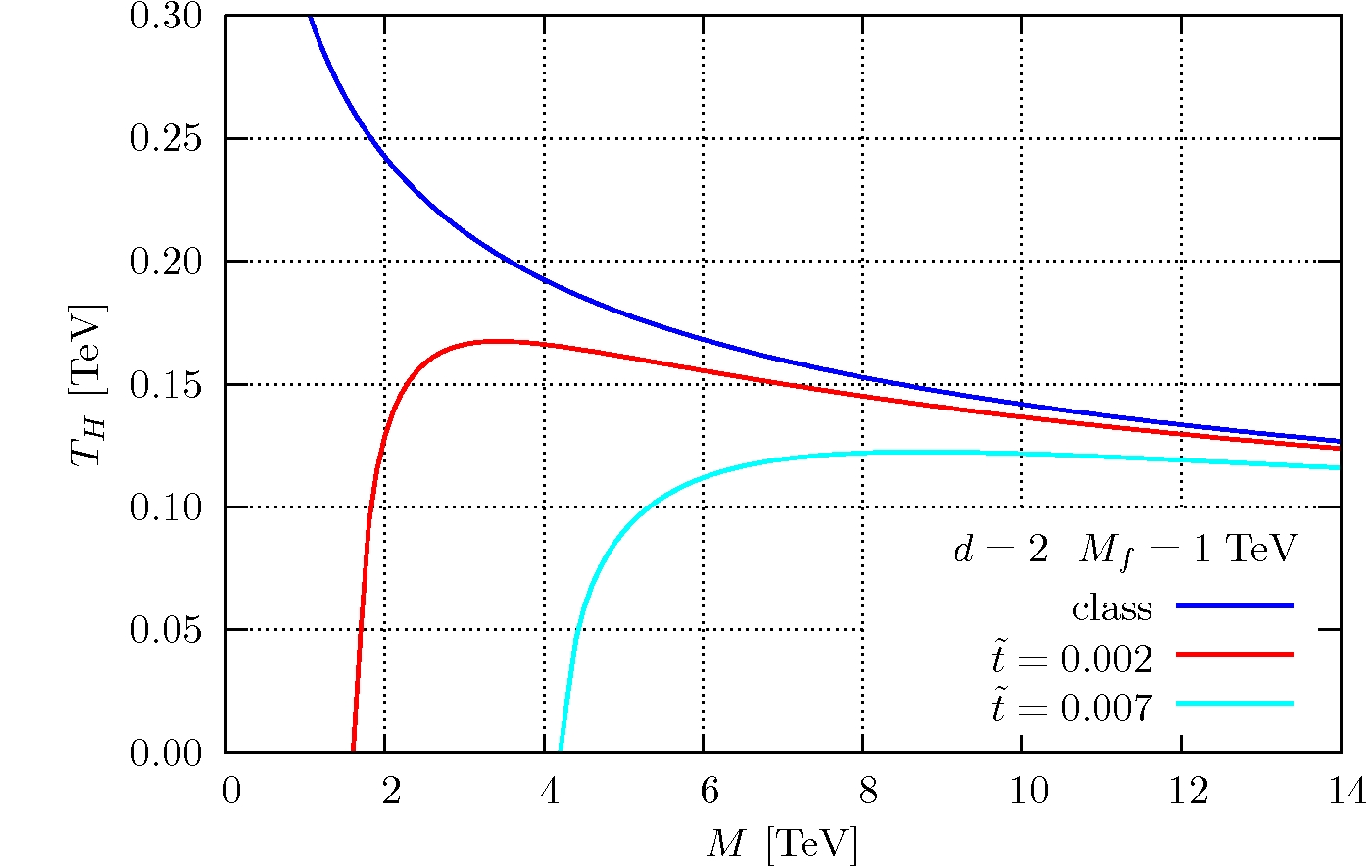}
\caption{Comparison of the standard and two
RG ($\tilde t=0.002$ and $\tilde t=0.007$, $M_f=1$\;TeV) temperature 
evolution of a black
hole with mass $M$.}
\label{fig:temp}
\end{figure}

The predicted thermal decay
of black holes due to Hawking radiation
can be considered a key testing stone for any suggested theory
of quantum gravity. 
Thus, black hole decay will now be studied in the presented theory
with large extra dimensions and RG improved black hole space-times.
The derivative of the radial function
at the event horizon will still be interpreted as the black hole temperature
\be
T_H=\left.\frac{1}{4 \pi}(\partial_r f(r))\right|_{r=Horizon} \ .
\ee
In a purely classical calculation, the temperature
rises for smaller BH masses, even up to the unphysical case
when the typical energy of single quantum emitted by the
black hole exceeds the total energy (mass) of the black hole. 
As it can be seen in figure\;(\ref{fig:temp}), the temperature
for the RG behaves like the standard Hawking temperature
for large masses $M\gg M_c$.
%
%%%%%%%%%% FIGURE %%%%%%%%%%%%%%%%%%
%%%%%%%%%%%%%%%%%%%%%%%%%%%%%%%%%%%%
But as soon as the BH mass approaches the critical mass
the RG corrected temperature is suppressed until at $M=M_c$
the temperature is zero. This object
with non zero mass $M_c$ but zero temperature is not emitting
any radiation and can thus be identified with a stable 
final BH state. Although the existence of such BH remnant states
can be motivated from different grounds
\cite{relics,Koch:2005ks,Vilkovisky:2005cz}
it is a natural outcome of the RG improved BH space-time.
This behavior solves the information loss problem and
the problem of unphysical (the total energy exceeding)
radiation for asymptotically slowly evolving black holes.

\begin{figure}[hbt]
\centering
%\centerline{\protect\vbox{\epsfig{file=Ivsomega.eps,
%width=0.6\textwidth}}}
\includegraphics[width=10cm]{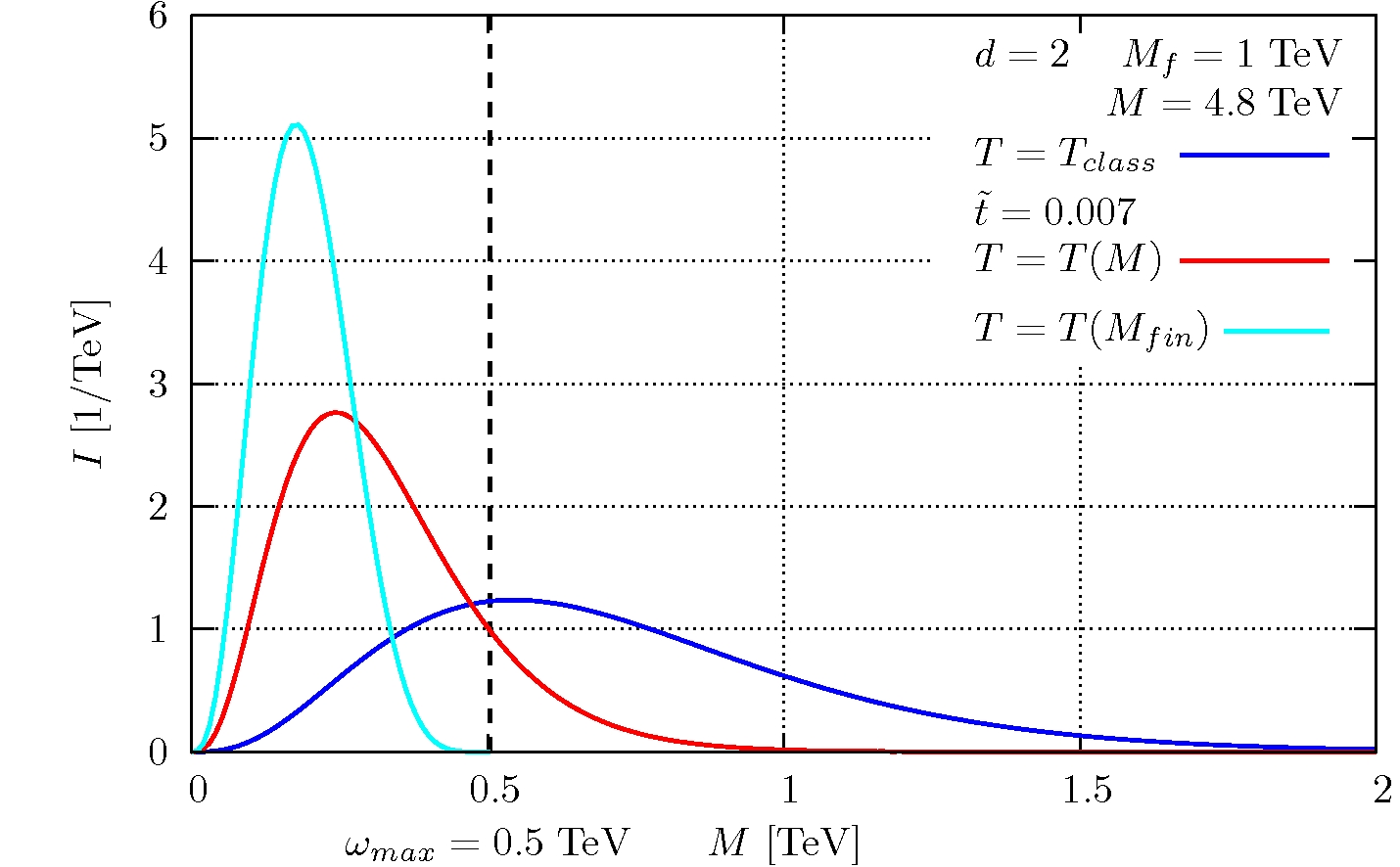}
\caption{Radiation spectrum for
RG ($\tilde t=0.007$, $M_f=1$\;TeV) temperature and a standard temperature
\eqref{eq:TH} for a black hole of
the mass $M=4.8$ TeV. The corresponding critical mass is $M_c=4.3$\;TeV.
Thus the maximally allowed $\omega$ 
for $m_\omega=0$ should be just below $0.5$\;TeV,
%0.47 TeV
which does not hold for the standard blue and the RG 
red curve, where $T_H=T_H(M)$ is assumed,
but which is true for the cyan curve where $T_H=T_H(M_{fin})$
is assumed. Here Boltzmann statistics ($s=0$) is used.}
\label{fig:spec}
\end{figure}

However, by looking at the thermal spectrum corresponding
to a given temperature one sees that one problems remains.
For thermal emission onto the four dimensional brane the
spectrum $I$ is given by
\be\label{eq:spec}
I(\omega,T_H)=N \frac{\omega^3}{\exp (\omega/T_H)+s} \quad, 
\ee 
where $N$ is a normalization factor and
$s$ is the factor corresponding to the spin
of the emitted particle 
(Fermi-Dirac: $s=1$, Boltzmann: $s=0$, Bose-Einstein: $s=-1$).
The energy (mass) of the black hole after a single
radiation process is then given by $M_{fin}=M-\omega$. 
Calculating the spectrum (\ref{eq:spec}) for a given black hole mass
$M>M_c$ and taking the temperature as the temperature
of the black hole before emission $T=T(M)$ one finds
that part of the previous problem persists:
The spectrum is non zero even for very large values
of $\omega$ (see fig.\;\eqref{fig:spec}), leading to the problem that
the BH mass after emission $M_{fin}=M-\omega$ still could be below
the critical mass or even negative.
The nature of this problem is similar to the anterior
problem of temperatures exceeding the total energy.

As simple mathematical solution of this problem
one can use the conservation
of energy and momentum. For
a black hole with mass $M$ that sits initially at rest and
emits a quantum with mass $m_\omega$ and energy $\omega$
the final mass is
\be
M_{fin}=\sqrt{M^2+m_\omega^2-2E_\omega M}\quad.
\ee
We propose to take
the RG-improved temperature as a function of
this final mass
\be
T_H=T_H(M_{fin})\quad,
\ee
 as opposed
to taking the dependence on the initial mass only $T_H=T_H(M)$.
Now, for any unphysical radiation like $\omega\ge M-M_c$, the temperature
in the exponential $T_H$ is zero and the spectral density (\ref{eq:spec})
vanishes. This cut off behavior is shown in figure\;(\ref{fig:spec}).
%%%%%%%%%% FIGURE %%%%%%%%%%%%%%%%%%
%%%%%%%%%%%%%%%%%%%%%%%%%%%%%%%%%%%%
The physical interpretation of this simple modification is:
First, one notices that for $\omega \ll (M-M_c)$ the modified
and the original spectrum agree.
Second, for $\omega \lesssim M-M_c$ the modification becomes important.
Further, the form $T_H=T_H(M_{fin})$ means that the 
black hole must know about the frequency $\omega$ of the emitted
quantum, already at the moment of emission, a behavior which
smells like violation of causality.
Nevertheless, in a thermodynamical approach this seems to be
simplest solution to the problem of overradiation.

The result of this section is that RG
(for a positive parameter $\tilde t$)
allows for a consistent description of the thermodynamic evolution
of black holes. It further predicts the formation of
a final stable black hole state with temperature $T_H=0$
and mass $M_c$.

%%%%%%%%%%%%%%%%%%%%%%%%%%%%%%%%%%%%%%%%%%%%%%%%%%%%%%%%%%%%%%%%%%%%%%%%%%%%5
\section{Phenomenology}
Several models with extra dimensions allow for
an effective Planck mass at the order of a few
TeV. The most exciting prediction of such models is the
production of mini black holes due to particle collisions
at the TeV scale. 
Already a simple implementation
of a running of the gravitational coupling has a significant
phenomenological impact on models with large extra
dimensions \cite{Hewett:2007st}.
Therefore, it is straight forward to study how the
predictions about mini black holes due to high energy particle collisions
change for the RG-improved black holes.
%\subsection{Black hole production}
The necessary condition for doing phenomenology with
such black hole is that they are produced at
all and that they are produced at sufficient rates. 
Therefore, we will leave the
analysis of the specific thermodynamical
properties or of the direct detection stable remnants 
to future studies and focus on the RG effects on 
black hole production.

\begin{figure}[bht]
\centering
%\centerline{\protect\vbox{\epsfig{file=areavsM_d2_Mf1_1.eps,width=0.6\textwidth
%}}}
\includegraphics[width=10cm]{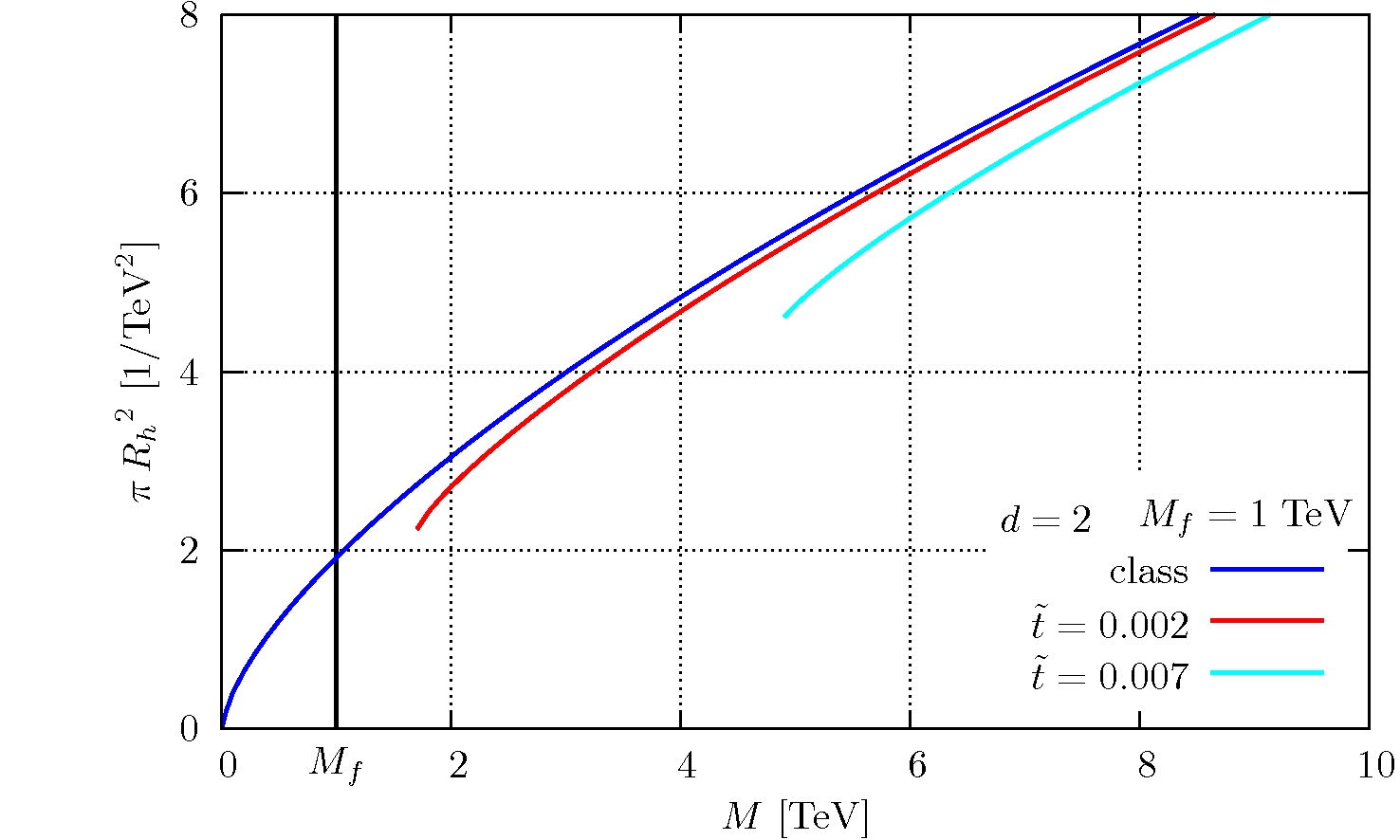}
\caption{Black hole cross sections for $d=2$ and $M_f=1$ TeV, 
varying $\tilde t$.}
\label{fig:cross-sect}
\end{figure}

The semi-classical cross section for the
production a black holes due to a particle collisions
with invariant energy $\sqrt{s}$ is given by
\be\label{crosssecalt}
\sigma(\sqrt{s})= \pi R_H^2 \theta(\sqrt{s}-M_{cut})\quad,
\ee
where $R_H$ is the Schwarzschild radius corresponding to the energy
$\sqrt{s}$. The production threshold
is usually associated with the higher dimensional Planck scale
$M_{cut}=M_f$.
At this point it should be mentioned that the possible
production of mini black holes at particle colliders 
does not imply any risk 
\cite{Giddings:2008gr,Giddings:2008pi,Koch:2008qq,Casadio:2009ri}.
This approximation of the cross section turned out to also be valid
in different approaches (for a discussion see 
\cite{Voloshin:2001fe,Jevicki:2002fq,Eardley:2002re,Rychkov:2004sf,Kang:2004yk,
Rizzo:2006uz}).
A first generalization is achieved by replacing the
classical Schwarzschild radius $R_H$ with the RG
improved Schwarzschild
radius $\tilde R_H$ \eqref{eq:imp:rh} (evaluated at the outer horizon).
The second generalization comes when replacing the heuristic
threshold mass $M_{cut}$ by the physical mass scale $M_c$. 
This leads to a physically intuitive threshold,
since only for $M>M_c$ an event horizon (and therefore
a black hole) exists.
Thus, taking renormalization group into account the
black hole cross section reads
\be\label{crosssec}
\tilde \sigma(\sqrt{s})= \pi \tilde R_H^2 \theta(\sqrt{s}-M_{c})\quad.
\ee
As one can see in figure\;(\ref{fig:cross-sect}),
for the production of very massive black holes $M\gg M_c$
the improved cross-section agrees with the semi-classical estimate.
Only when $M$ is slightly higher than $M_c$,
the numerical values start to differ significantly.
The most drastic difference to the semi-classical estimate appears when $M=M_c$,
since this defines the new threshold for black hole production.
One sees that this threshold $M_c$ can differ largely from
the ad hoc threshold $M_f$.
Since the remnant mass depends strongly on the RG
parameter $\tilde t$, the black hole
threshold also
depends strongly on $\tilde t$. 
%
%%%%%%%%%% FIGURE %%%%%%%%%%%%%%%%%%
%%%%%%%%%%%%%%%%%%%%%%%%%%%%%%%%%%%%
An anologous result was obtained in \cite{Litim:2008xxx}.
Please note that if one just replaces
$M_f$ by $\tilde M_f(k)$ without taking the 
modification of $f(r)$ into account one finds that
the cross section for large masses $M\gg M_f$ 
is damped to zero \cite{Koch:2008zzb}.
As it can be seen in figure\;(\ref{fig:cross-sect})
the RG-improved result does approach the semi-classical estimate
for $M\gg M_c$ and
therefore the damping behavior was just an
artefact of not taking the modification of $f(r)$ into account.

\begin{figure}[bht]
\centering
%\centerline{\protect\vbox{\epsfig{file=d2ttvsMf.eps,
%width=0.6\textwidth}}}
\includegraphics[width=10cm]{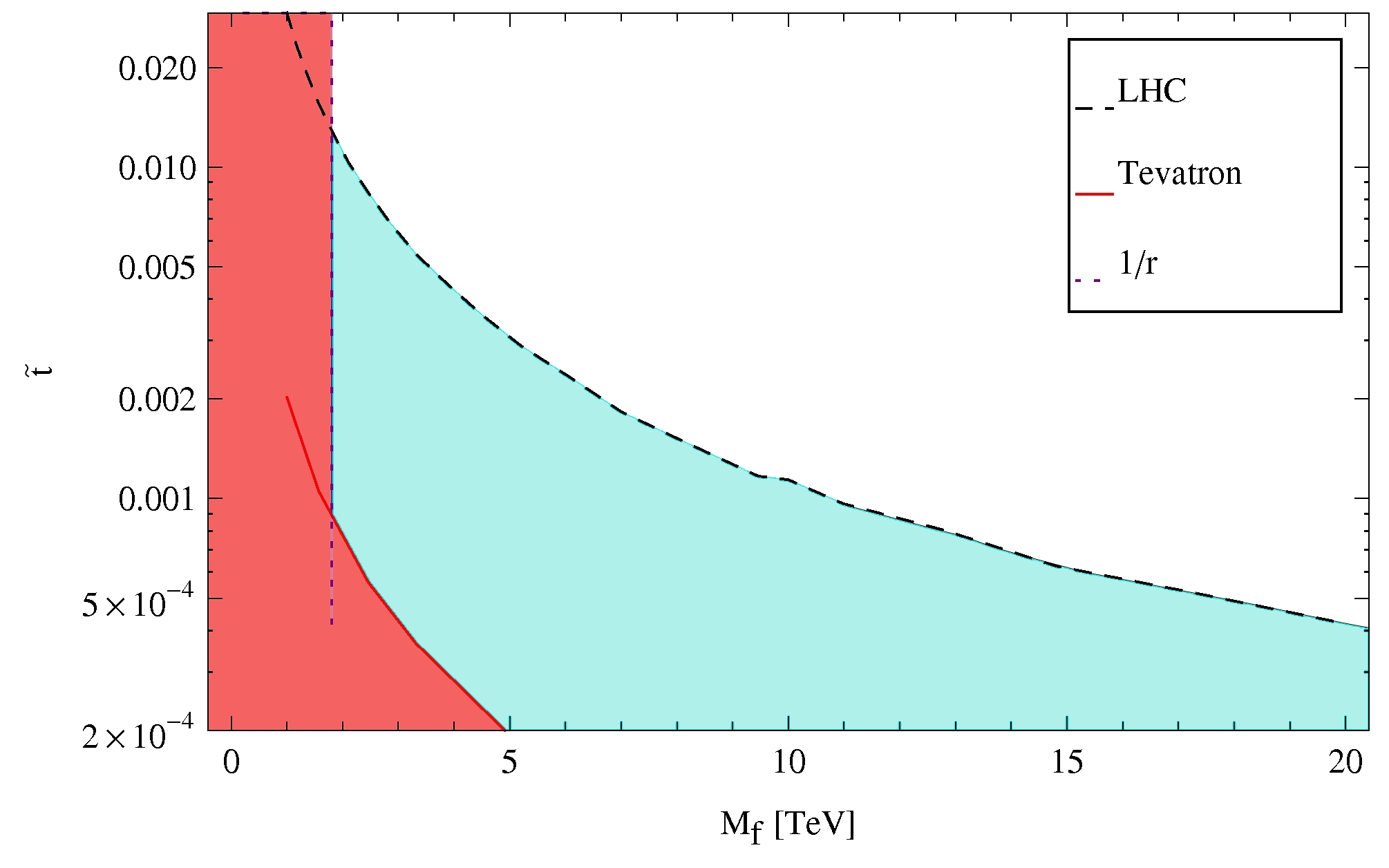}
\caption{LHC ($\sqrt{s}=14$\,TeV, $L_{LHC}= 100$ $\rm{fb}^{-1}$) discovery
potential (turquoise region) in the
parameterspace $\tilde t$
and $M_f$ for $d=2$.
The Tevatron line is calculated with equation (\ref{partint}) for 
$\sqrt{s}=1.8$\,TeV and $L_{Tev}= 10$ $\rm{fb}^{-1}$.
For $d=2$ the relation (\ref{eq:int:relation}) in combination with
%$V_d=(2Pi R)^d
the experimental limit  
on the compactification radius $R\le 0.1\;$mm \cite{Kapner:2006si}
gives the vertical line at $M_f\approx 1.8\;TeV$.
Thus the red region is already excluded and the white region is not accessable
by the LHC experiment.}
\label{fig:parspaced2}
\end{figure}

Now, an estimate of the LHC discovery potential on the parameter
space $d$, $M_f$, and $\tilde t$ will be given. 
Clearly, an upper limit of the accessable production
threshold $M_c$ will be given by the maximal LHC center of mass
energy of $14$ TeV. While an upper limit on the accessable cross-section area 
$\pi \tilde R_H^2$ will be given by the first mileston of the integrated LHC
luminosity $L_{LHC}\sim 100$ $\rm{fb}^{-1}$.
The same estimate can be made for the Tevatron with
$M_c<1.8\;\rm{TeV}$ and $\pi \tilde R_H^2<L_{Tev}=10 \rm{fb}^{-1}$).
Combining the estimates already gives a 
rough estimates on lower limits
on the LHC discovery potential in this model.
For a more quantitative estimate of the discovery potential
the luminosities ($L_{LHC}, \,L_{Tev}$)
are compared to
integrated partonic cross section
\be\label{partint}
\sigma(\sqrt{s})=\sum_{i,j}^{partons}
\int_0^1 dx \int_0^1 dy \,f_i(x,q) f_j(y,q)
\tilde \sigma (\sqrt{\hat s})\quad,
\ee
by using the parton distribution functions $f_i$ \cite{Alekhin:2002fv}.
Here, the center of mass energy in the partonic
reference frame is given by $\sqrt{\hat s}=\sqrt{(x y) s }$.
For $d=2$ the result of this scan over the parameter space is given in
figure\;(\ref{fig:parspaced2}), showing that the parameter $\tilde t$
has to be tuned to small values in order to
be testable at one of the LHC experiments.
On the other hand one sees that given a small value of $\tilde t$,
the LHC might produce mini black holes even
if the fixed point mass $M_f$ exceeds the center of mass energy of 14\;TeV.
For each additional number of extra dimensions, the LHC and Tevatron
lines are shifted about one order magnitude towards smaller values
of $\tilde t$.
%
%%%%%%%%%% FIGURE %%%%%%%%%%%%%%%%%%
%%%%%%%%%%%%%%%%%%%%%%%%%%%%%%%%%%%%
%This statement becomes even
%stronger for larger number of extra dimensions $d>2$.

\section{Summary and conclusion}
We studied the impact of an effective running
of the gravitational coupling on Schwarzschild black
hole space-times for the case of large extra dimensions.
The analysis of the black hole evaporation process
revealed that this approach allows to solve
the problem of overradiation for the whole process even for
black hole masses comparable to the fundamental mass scale $M\sim M_f$,
which was not possible in the standard description 
(see fig.\;\ref{fig:temp},\;\ref{fig:spec}).
Further, it was found that this evaporation ends in
a stable final state with mass $M_c$.
Finally the impact of this approach on 
the predicted black hole production at the large hadron
collider is calculated and an estimate
of the possible discovery potential is given (see fig.\;\ref{fig:parspaced2}).\\

T.\,B. thanks M. Bleicher for supervision and ITP Frankfurt for accommodation.
Many thanks also to D. Litim and C. Rahmede for very helpful comments
and remarks. 
B.\,K. was funded by Conicyt-PBCT grant PSD73.

\end{document}